\documentclass[manuscript]{aastex}

\def\lapp{\ifmmode\stackrel{<}{_{\sim}}\else$\stackrel{<}{_{\sim}}$\fi}
\def\gapp{\ifmmode\stackrel{>}{_{\sim}}\else$\stackrel{<}{_{\sim}}$\fi}
\def\farcm{\hbox{$.\mkern-4mu^\prime$}}

\shorttitle{A Shapiro delay detection in [PSR\,J1910-5959A+co].}
\shortauthors{Corongiu et al.}

\begin{document}

\title{A Shapiro delay detection in the binary system hosting
the millisecond pulsar PSR\,J1910-5959A.}

\author{A. Corongiu\altaffilmark{1},
M. Burgay\altaffilmark{1},
A. Possenti\altaffilmark{1},
F. Camilo\altaffilmark{2},
N. D'Amico\altaffilmark{1,3},
A.G. Lyne\altaffilmark{4},
R.N. Manchester\altaffilmark{5},
J.M. Sarkissian\altaffilmark{6},
M. Bailes\altaffilmark{7},
S. Johnston\altaffilmark{5},
M. Kramer\altaffilmark{4,8},
W. van Straten\altaffilmark{7}}
\altaffiltext{1}{INAF - Osservatorio Astronomico di Cagliari, 
              Loc. Poggio dei Pini, Strada 54, 09012 Capoterra (CA), Italy}
\altaffiltext{2}{Columbia Astrophysics Laboratory, Columbia University,
              550 West, 120th Street, New York, NY 10027}
\altaffiltext{3}{Universit\`{a} degli Studi di Cagliari, Dip. di Fisica, 
	      S.S. Monserrato-Sestu km 0,700, 09042 Monserrato (CA), Italy}
\altaffiltext{4}{Jodrell Bank Centre for Astrophysics, School of Physics
              and Astronomy,
              University of Manchester, Manchester M13 9PL, UK}
\altaffiltext{5}{CSIRO Astronomy and Space Science, Australia Telescope National
              Facility, PO Box 76, Epping, NSW 1710, Australia}
\altaffiltext{6}{Australia Telescope National Facility, CSIRO, Parkes
              Observatory, P.O. Box 276, Parkes, New South Wales 2870,
              Australia}
\altaffiltext{7}{Centre for Astrophysics and Supercomputing, Swinburne
              University of Technology, Post Office Box 218 Hawthorn, VIC 3122,
              Australia}
\altaffiltext{8}{Max-Planck-Institut fur Radioastronomie, Auf dem H\"{u}gel 69,
              53121 Bonn, Germany}

\begin{abstract}
PSR\,J1910-5959A is a binary pulsar with a helium white dwarf
companion located about 6 arcmin from the center of the globular
cluster NGC\,6752. Based on 12 years of observations at the Parkes
radio telescope, the relativistic Shapiro delay has been detected in
this system. We obtain a companion mass $M_C\,=\,0.180\pm0.018M_\odot$
(1$\sigma$) implying that the pulsar mass lies in the range
$1.1\,M_\odot\,\leq\,M_P\,\leq\,1.5\,M_\odot$. We compare our results
with previous optical determinations of the companion mass, and
examine prospects for using this new measurement for calibrating the
mass$-$radius relation for helium white dwarfs and for investigating
their evolution in a pulsar binary system. Finally we examine the set
of binary systems hosting a millisecond pulsar and a low-mass helium
white dwarf for which the mass of both stars has been measured. We
confirm that the correlation between the companion mass and the
orbital period predicted by \citeauthor{ts99a} reproduces the observed
values but find that the predicted $M_P-P_B$ correlation
over-estimates the neutron-star mass by about $0.5M_\odot$ in the
orbital period range covered by the observations. Moreover, a few
systems do not obey the observed $M_P-P_B$ correlation. We discuss
these results in the framework of the mechanisms that inhibit the
accretion of matter by a neutron star during its evolution in a
low-mass X-ray binary.
\end{abstract}


\section{INTRODUCTION}

PSR\,J1910-5959A (henceforth PSRA) is a millisecond pulsar (MSP)
discovered in the Parkes Globular Cluster Pulsar Survey
(\citealt{dlm+01}, hereafter paper\,I). It has a spin period of
3.27\,ms, and it orbits around a low-mass companion with
$M_C\,\geq\,0.2\,M_\odot$, assuming a pulsar mass
$M_P\,=\,1.4\,M_\odot$, with an orbital period $P_B\,=\,0.84$\,days.
Because of its unusual position with respect to the core of the
globular cluster NGC\,6752 (its angular distance from the cluster
center is $\theta_\perp\,=\,6\farcm4$, corresponding to 3.3 half-mass
radii; paper\,I), it has been questioned whether this object is a
member of the cluster \citep{bvkv06}. In fact, of all
cluster-associated MSPs, PSRA has the largest angular distance from
the core of the hosting cluster.\footnote{Intriguingly PSR\,J1910-5959C
  in the same cluster ranks second in this list, being located at
  $\theta_\perp\,=\,2\farcm7$, corresponding to 1.4 half-mass radii,
  from the core of the cluster (paper\,I).}

The binary companion of PSRA has been recognised as a helium white
dwarf star (HeWD) by \citet{fps+03} and independently by
\citet{bvkh03}, via observations taken with the ESO Very Large
Telescope (VLT) and the Hubble Space Telescope (HST). The values
obtained by these authors for the mass of the binary companion of PSRA
are in agreement ($M_{\rm co}\simeq\,0.17-0.20\,{\rm
M_\odot}$).  Both authors also agree in concluding that the
photometric properties of this star are consistent with the hypothesis
that PSRA is associated with the globular cluster NGC\,6752.

The radial velocity curve of the companion of PSRA has been
investigated by \citet{cfp+06} and \citet{bvkv06}, through
phase-resolved spectroscopy in the optical band with HST
observations. The resulting values for the mass ratio of the binary
system, for the mass of the optical companion and for the orbital
inclination were also compatible with each other, although
\citet{bvkv06} obtained more stringent constraints (see
\S\ref{subsec:impl} for details).  \citet{bvkv06} carefully discussed
the association of this binary system with NGC\,6752, and indicated a
preference for non-association, though their arguments are not
conclusive because of the large uncertainties on the parameters
that would allow one to discriminate between the two scenarios (see
\S\ref{subsec:impl} for details).

A refined ephemeris for PSRA was presented by \citet{cpl+06}
(hereafter paper\,II), where a low eccentricity model was applied
to describe the pulsar's orbit. The measured eccentricity was
$e\sim3\times10^{-6}$ at a longitude of periastron
$\omega\sim\,90^\circ$. Paper\,II discussed the possibility
that such eccentricity was indeed a misinterpreted Shapiro delay.
Nevertheless, paper II maintained a conservative approach to this
problem and deferred the revision of this issue until a significantly
longer time span of timing data became available.

Correlations between the orbital period, the pulsar mass and the
companion mass have been obtained by \citeauthor{ts99a}
(\citeyear{ts99a}, hereafter TS99), from their numerical calculations
on the non-conservative evolution of close binary systems with
low-mass donor stars ($M_D\,<\,2\,M_\odot$), a 1.3\,$M_\odot$
accreting neutron star and an orbital period $P_B\,\geq\,2$\,days at
the beginning of their low-mass X-ray binary (LMXB) phase.  These
systems have diverging orbital separation and are the progenitors of
low-mass binary millisecond pulsars (LMBMSP, \citealt{bv91}). TS99
obtained a positive correlation between the orbital period and the
companion mass and a negative correlation between the orbital period
and the pulsar mass.  The predicted $M_C-P_B$ correlation has been
observed by \citet{vbjj05} and \citet{l08}. TS99 remarked on the
discrepancy between the predicted $M_P-P_B$ correlation and the
available data. However in most cases the determinations of the $M_C$
and $M_P$ were based on models for the companion atmosphere or on
statistical hypothesis for the orbital inclination, often leading to
large uncertainties. The observation of a Shapiro delay 
allows one to obtain a direct measurement of $M_C$ and $M_P$ and
gives a strong test for the aforementioned correlations.

In this paper we present timing data for PSRA collected over 12 years
at the Parkes radio telescope. This data span is twice than that used
in paper\,II and allowed us to get better constraints on the
rotational, positional and orbital parameters for PSRA, and to detect
the Shapiro delay in this system. The paper is organized as
follows. In section \S\ref{sec:obs} we briefly describe the observing
methods and present the timing solution based on the available data
span, in \S\ref{sec:shapiro} we describe the method for measuring the
Shapiro delay, present our determination of the companion mass and
discuss our results and their implications. In \S\ref{sec:corr} we
summarize the measurements for the masses of MSPs and their HeWD
companions, focussing on those binaries for which $M_C$ and $M_P$ have
been obtained from the detection of the Shapiro delay effect. We then
compare these data to the predictions of TS99 and discuss the
results. In \S\ref{sec:summary} we briefly summarise our work.

\section{OBSERVATIONS AND GENERAL TIMING RESULTS}
\label{sec:obs}

\begin{figure} \centering
\includegraphics[angle=-90,width=15cm]{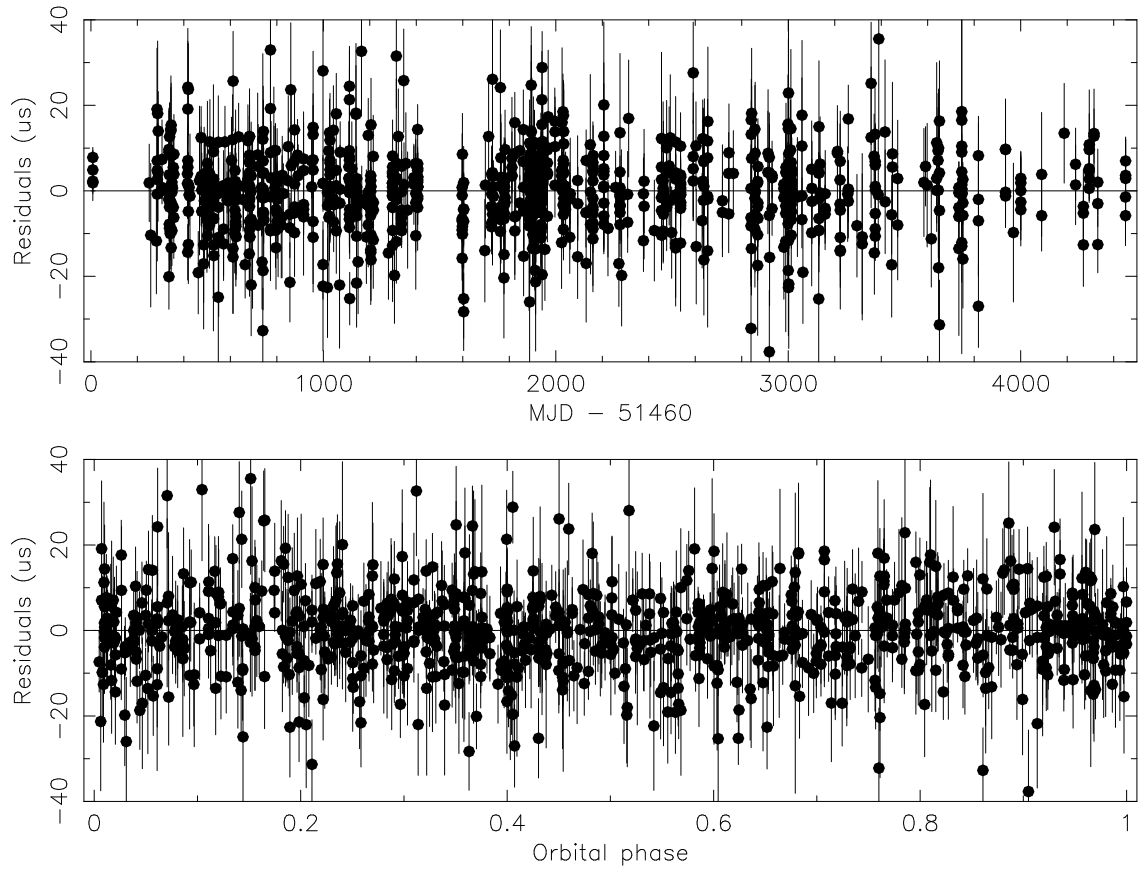}
\caption{Timing residuals versus Modified Julian Date (MJD, upper panel)
and orbital phase (lower panel) for the timing solution presented in this
work.
\label{fig:residuals}}
\end{figure}

Regular pulsar timing observations of PSRA in NGC\,6752 have been
carried out since September 2000 with the Parkes 64\,m radio telescope
at a central frequency of 1390\,MHz, using the central beam of the
20-cm multibeam receiver \citep{swb+96} or the H-OH receiver. The
hardware system is the same as that used in the discovery observations
(paper\,I). The effects of interstellar dispersion are minimised by
using a filterbank having 512$\times$0.5\,MHz frequency channels for
each polarisation.  After detection, the signals from individual
channels are added in polarisation pairs, integrated, 1-bit digitised
every 80\,$\mu$s (125\,$\mu$s in earlier observations), and recorded
to magnetic tape for off-line analysis.  We synchronously folded the
data at the pulsar period using the program {\tt DSPSR} \citep{vb11}
with a subintegration time of 1\,min and dedispersion in groups of 64
channels to give eight frequency sub-bands. We visually inspected each
file to extract the maximum number of pulse times of arrivals (ToAs),
depending on integration time and interstellar scintillation which
causes large fluctuations in the detected flux density. We considered
a ToA reliable if the pulse profile had a signal-to-noise ratio
greater than 5 and the pulse was visible.  This approach gave us $\sim
1000$ ToAs at different frequencies within the band of the receivers
and consequently allowed us to fit for the dispersion measure and its
first derivative. We calculated the ToAs by fitting a template profile
to the observed mean pulse profiles and we analysed them using the
program {\tt TEMPO2} \citep{hem06} with the DE405 solar system
ephemeris and the {\tt TT(TAI)} reference
timescale. Table\,\ref{TabTim} summarises the best fit values and
$1\sigma$ uncertainties from {\tt TEMPO2} (everywhere in the paper
all reported measures are quoted with their $1\sigma$ uncertainty,
and all reported ranges are at the same confidence level). The
timing residuals for the fit which have a $\chi^2$ of 989.3 (with 989
degrees of freedom) are shown in Fig.\,\ref{fig:residuals}, against
MJD in the upper panel and against orbital phase in the lower
panel. We determined the values and the uncertainties for the Shapiro
delay parameters through a Bayesian analysis (see
\S\ref{sec:shapiro}).

The new positional and rotational parameters at the reference epoch
are all consistent with those reported in paper\,I and paper\,II at a
confidence level of $2\sigma$, with the only exception being the
proper motion, whose components in right ascension and declination are
consistent at the 4$\sigma$ and 6$\sigma$ level respectively. The
value for the first derivative of the dispersion measure (DM1) is of
the same order of magnitude of similar determinations for other
MSPs. The consistency of the orbital parameters will be discussed in
\S\ref{sec:shapiro}.

\section{THE SHAPIRO DELAY MEASUREMENT}
\label{sec:shapiro}

\subsection{Method}
\label{subsec:method}

The orbital solutions presented and discussed in papers I and II were
based on the ELL1 binary model, which is suitable for systems with
small orbital eccentricities (Wex 1998 unpublished work; see
\citealt{lcw+01}). In paper\,II an orbital eccentricity
$e\,=\,(3.4\pm0.6)\times10^{-6}$ was reported, at a
longitude of periastron $\omega\,=\,90^\circ\pm10^\circ$. The values
obtained for $e$ and in particular for $\omega$ were already commented
as a possible misinterpretation of a Shapiro delay (paper\,II). By
applying the same binary model of paper\,II to the longer data span
presented in this work, we obtain $e\,=\,(3.5\pm0.5)\times10^{-6}$ at
a longitude of periastron $\omega\,=\,95^\circ\pm7^\circ$. Since the
uncertainties on these two parameters did not decrease as expected for
the case of a pure Keplerian ELL1 model we carefully reinvestigated
the possible presence of a Shapiro Delay.

\begin{figure} \centering
\includegraphics[angle=-90,width=15cm]{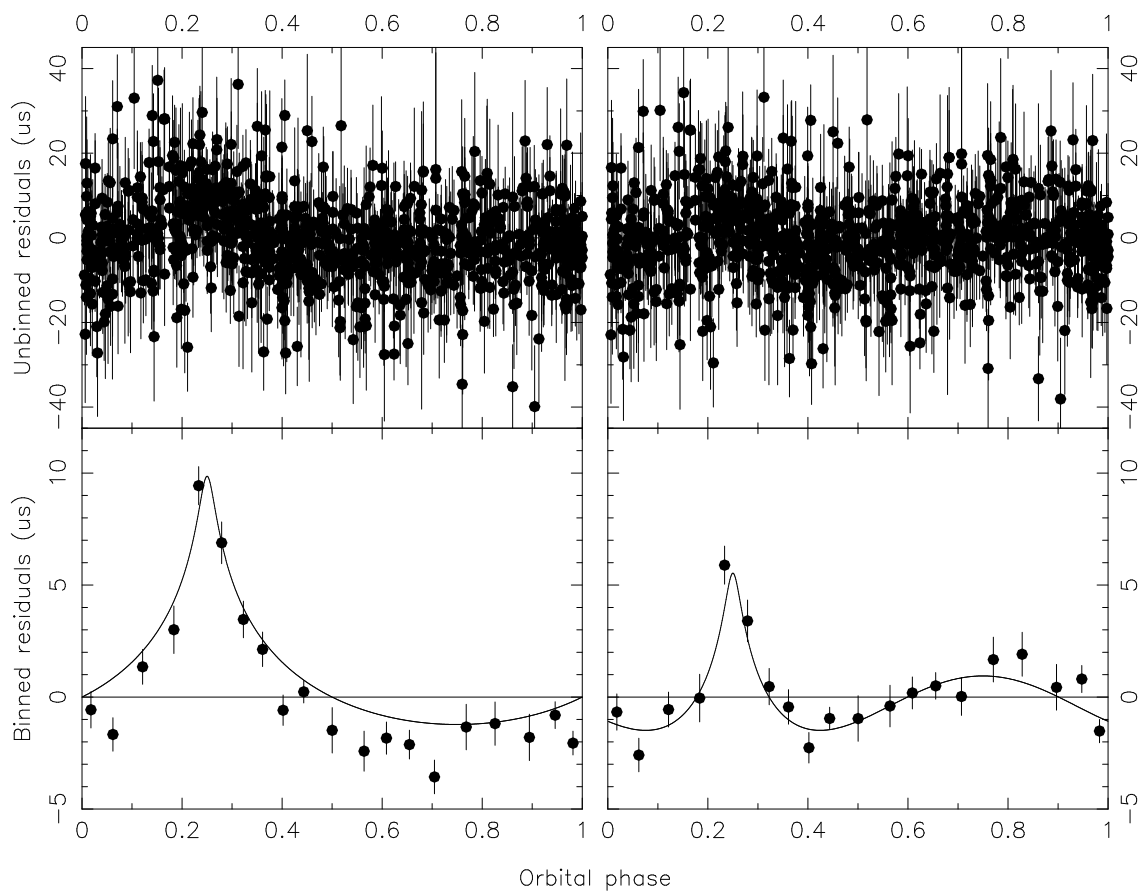}
\caption{Timing residuals versus orbital phase. Upper panels display
unbinned residuals, while lower panels display residuals after being
binned in 20 orbital phase bins. The left panels are related to the
attempt of detecting the first harmonic of the Shapiro delay, while
the right panels concern the third harmonic. The solid lines in the lower
panels are the theoretical curves for the respective harmonics.
\label{fig:shapiro}}
\end{figure}

We used the DD binary model (\citealt{dd85}; \citealt{dd86}) and
included the two parameters that describe the Shapiro delay, namely
the range $r$ and the shape $s$. These two parameters are directly
related to the companion mass $M_C$ and the orbital inclination $i$
through the simple relations $r\,=\,T_\odot M_C$
($T_\odot\,=\,GM_\odot/c^3$, where $G$ is Newton's universal
gravitation constant and $c$ is the speed of light) and $s\,=\,\sin i$
(see, e.g., \citealt{lk05}).  Fig.\,\ref{fig:shapiro} displays our
attempts to detect the first and third harmonic of the Shapiro delay
in our timing residuals. We recall (e.g. \citealt{lk05}) that the
first harmonic is obtained by fitting for all parameters, including
the ones describing the Shapiro delay, then resetting $M_C\,=\,0.0$
and $\sin i\,=\,1.0$ while keeping all other parameters to their best
fit values, and finally by plotting the resulting timing
residuals. The third harmonic is obtained by fitting for all
parameters but the Shapiro delay ones, which remain set to the values
$M_C\,=\,0.0$ and $\sin i\,=\,1.0$, then plotting the residuals for
this best fit solution obtained in this way. While the first harmonic
is still consistent with the response for an elliptic orbit, the third
harmonic is a unique signature of the Shapiro delay. In the upper
panels of Fig.\,\ref{fig:shapiro} timing residuals show some evidence
of the harmonic structure, while in lower panels, where residuals have
been rebinned in 20 orbital phase bins, the two harmonics are clearly
recognisable. As a further comparison, in the lower panels we plot
(solid lines) the theoretical curves for the Shapiro delay with which
our rebinned residuals are in satisfactory agreement.

As an additional sanity check we have checked the consistency of the
three Keplerian parameters that are common to the two different
orbital models presented in paper\,II and in this work. The values for
$P_B$ are fully consistent to each other, and the same is true
comparing $T_{ASC}$ in paper\,II and $T_0$ here. We recall that in the
case of zero eccentricity orbits, the periastron is defined as
coincident to the ascending node. The only parameter for which we now
find an inconsistency at a level of about 10$\sigma$ with respect to
paper\,II is the orbit projected semi-mayor axis $a_P\sin i$. However
because of the change in the adopted binary model (from ELL1 to DD)
this parameter cannot be directly compared.  The quantity to be
compared between is $a_P\sin i/(1+e)$. From paper\,II we obtain
$a_P\sin i/(1+e)\,=\,1.2060425\pm0.0000007$, while in this work it
results $a_P\sin i\,=\,1.2060420\pm0.0000008$, as the orbital
eccentricity has been now set to zero. The two values are then in full
agreement.

We obtained the values for the companion mass and the inclination angle
through a Bayesian analysis on our timing residuals. The
method is described in \citet{sna+02}. We chose the a priori probability
density functions as follows:

\begin{itemize}
\item[a)]{A flat distribution for $\cos i$ in the range
$0\,\le\,\cos i\,\le\,1$, as we assumed a randomly oriented
orbit in space}
\item[b)]{A flat distribution for $M_C$ in the range
$0.0\,M_\odot\,\le\,M_C\,\le\,0.5\,M_\odot$, chosen for consistency
with the values of this parameter resulting from the study
of the radial velocity curve of the companion (\citealt{bvkv06};
\citealt{cfp+06})}
\end{itemize}

\begin{figure} \centering
\includegraphics[angle=-90,width=15cm]{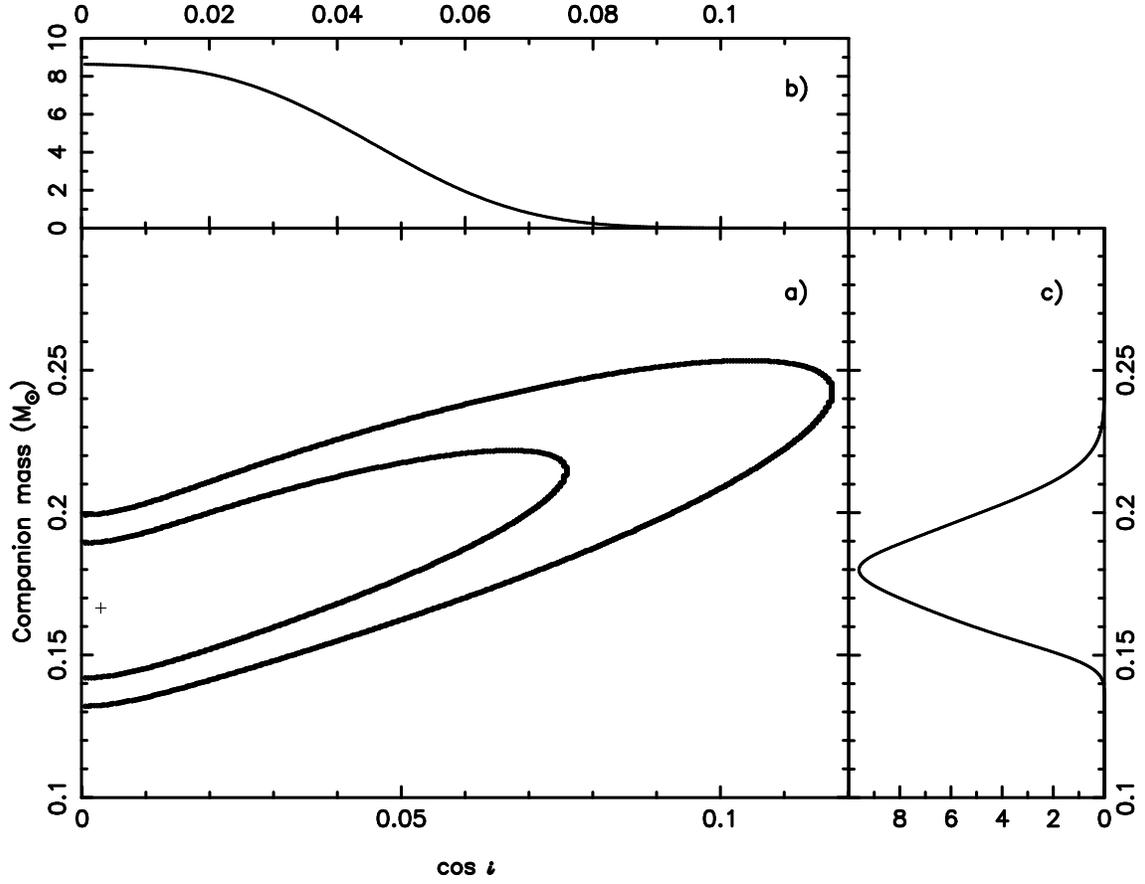}
\caption{Panel a) $\chi^2$ map in the $\cos i-M_C$ space. The cross is
located at the point of minimum $\chi^2$. The two solid lines
correspond to the $1\sigma$ (inner line) and $2\sigma$ (outer line)
regions. Panel b) Posterior probability density function for $\cos i$.
Panel c) Posterior probability density function for $M_C$.
\label{fig:chimap}}
\end{figure}

Fig.\,\ref{fig:chimap} displays in panel a) the $\Delta\chi^2$ map
obtained after calculating the $\chi^2$ values in a 2048x2048 grid for
the parameters of our interest. In panels b) and c) the posterior
probability density functions for $\cos i$ and $M_C$
respectively are reported. From this analysis we derived a lower
limit for the orbital inclination $\sin i\,\geq\,0.9994$,
i.e. $i\,\geq\,88^\circ$, and the value for the companion
mass $M_C\,=\,0.180\pm0.018\,M_\odot$.

\subsection{Implications}
\label{subsec:impl}

Under the hypothesis that PSRA is associated with NGC\,6752,
the measurement reported in \S\ref{subsec:method} is the second of
this kind for a binary pulsar in a globular cluster after
PSR\,J1807-2459B in NGC\,6544 \citep{lfrj12}.  PSR\,J1807-2459B is a
millisecond pulsar with a spin period $P_0\,=\,4.19$\,ms in a highly
eccentric orbit $e\,=\,0.747$. The nature of the companion is not yet
determined, however its high mass $M_C\,=\,1.21\,M_\odot$
points towards a neutron star or a massive white dwarf companion. Thus
PSRA represents the first case of a low-mass binary pulsar
in a globular cluster for which the Shapiro delay has been detected.

Our determination of the companion mass,
$M_C\,=\,0.180\pm0.018\,M_\odot$, is solely based on the relativistic
Shapiro delay and so we did not assume any a priori hypothesis on
the nature of the companion or the modelling of its structure.  In this
respect the full agreement with the values for $M_C$ obtained using
observations in the optical band provides independent support to
the previously suggested interpretation of the nature of the binary
companion of PSRA, i.e., that it is a low-mass helium-core white
dwarf (\citealt{bvkh03}, \citealt{bvkv06}, \citealt{fps+03},
\citealt{cfp+06}). This also gives support to the
physical modelling (i.e. the relationship between mass, radius,
temperature and surface gravity of this category of white dwarfs)
invoked for deriving the parameters of PSRA from the
photometric and spectroscopic data only.

\citet{bvkv06} attempted to use a comparison between the observed
parameters and the models mentioned above in order to decide whether
or not this binary system belongs to the globular cluster
NGC\,6752. Unfortunately the uncertainties in the measurements of the
companion parameters prevented them from drawing a firm
conclusion. Since our uncertainty on $M_C$ is similar to that of
\citet{bvkv06}, our result does not help in addressing this
point. Further support for the association of PSRA to NGC\,6752 is
expected from the measurement of proper motion for the three isolated
pulsars in the core of NGC\,6752\footnote{They certainly belong to
  NGC\,6752 since at least two of them show the large $|\dot P|$
  attributed to the gravitational pull of the
  cluster. Their proper motion has not been measured yet, since they
  are weaker sources.} and comparison with the
already accurately determined proper motion for PSRA (paper\,II).

Assuming that PSRA is associated with NGC\,6752, the timing
determination of the companion mass opens the possibility of using the
system for testing the various theoretical mass-radius relations
proposed for helium white dwarfs.  The large uncertainties in the
determination of the masses and of the distances of the white dwarfs
via optical observations make this difficult. This problem is
particularly important for the low-mass white dwarfs, for which
accurate determinations of the mass and the radius are rare. In this
case we have independent measurements for the mass of the companion of
PSRA and its distance, which is equal to the cluster's one under our
hypothesis. From optical data \citet{bvkv06} derive a companion
radius, namely $R_C\,=\,0.058\pm0.004\,R_\odot$ if PSRA belongs to
NGC\,6752.  We note that the mass-radius relations predicted by both
the models of \citet{rsab02} and \citet{sarb02} nicely agree with the
case of a white dwarf of $R_C\,=\,0.058\pm0.004\,R_\odot$ and
$M_C\,=\,0.180\pm0.018\,M_\odot$. The two models differ in the
progenitor's metallicity, namely the \citet{rsab02} describes low-mass
white dwarfs with solar metallicity while the \citet{sarb02} model is
suitable for metallicities comparable to the one of NGC\,6752. The
model by \citet{dsbh98} computes the white dwarf radius only for
masses $M_{WD}\,\geq\,0.180\,M_\odot$. It cannot be yet discarded (see
for reference fig.\,6 in \citealt{bvkv06}).

Given a typical $\sim$8\% uncertainty in the radius determination from
optical observations, in order to discriminate between various models
for the mass-radius relations, an uncertainty on $M_C$ of
0.002\,$M_\odot$ is typically required. Unfortunately several decades
are required to achieve this goal with a 64-m class telescope; such
high precision could be obtained in about one decade by observing PSRA
with the Square Kilometer Array.

A determination of the mass $M_P$ for PSRA can be obtained from our
Shapiro delay measurement.  From the lower limit  in the orbital
inclination and the pulsar mass function, we can obtain a firm range
for $M_P$, $1.1\,M_\odot\,\leq\,M_P\,\leq\,1.5\,M_\odot$, from which
we deduce the total mass for the system lying in the range
$1.26\,M_\odot\,\leq\,M_{\rm tot}\,\leq\,1.70\,M_\odot$. If we instead
use the mass ratio determined by combining the pulsar mass function
from pulsar timing and the companion's mass function from the studies
on the companion's velocity curve by \citet{bvkv06} and
\citet{cfp+06}, we can slightly improve our results. \citet{bvkv06}
report $M_P/M_C\,=\,7.36\pm0.25$, from which we now obtain $M_{\rm
tot}\,=\,1.50\pm0.14\,M_\odot$ and $M_P\,=\,1.32\pm0.14\,M_\odot$,
while \citet{cfp+06} report $M_P/M_C\,=\,7.49\pm0.68$, from which we
now obtain $M_{\rm tot}\,=\,1.53\pm0.18\,M_\odot$ and
$M_P\,=\,1.35\pm0.18\,M_\odot$. The weighted average of these two
values yields $M_P\,=\,1.33\pm0.11\,M_\odot$.

It is worth mentioning that the value for the mass of PSRA is close to
the lower edge of the mass range for binary millisecond pulsars
located in the Galactic disk whose companion is a HeWD star. Since in
these systems the neutron star has experienced only one recycling
phase, thus leaving the original neutron star mass almost unchanged,
the value for the mass of PSRA suggests that this pulsar underwent only
one recycling stage. If we combine this statement with the hypothesis
that PSRA belongs to the cluster, we can speculate that the ejection
of this system towards the cluster's outskirts (\citealt{cpg02};
\citealt{cmp03}) most likely happened in an early stage of the cluster
evolution, preventing this binary from experiencing further exchange
interactions with the stars in the crowded environment of the
cluster's core. This appears to be at odds with the predictions of the
models which suggest that the ejection occurred in the last
$\sim1$\,Gyr \citep{cpg02}.

\section{CORRELATIONS AMONG THE ORBITAL PERIOD, THE
PULSAR AND THE COMPANION MASS FOR LOW MASS BINARY
PULSARS}
\label{sec:corr}

TS99 presented detailed numerical calculations on the non-conservative
evolution of close binary systems with a low-mass
($M_D\,<\,2.0\,M_\odot$) donor star, a $1.3\,M_\odot$ accreting
neutron star and an orbital period $P_B\geq2$\,days at the beginning
of the LMXB phase.  The evolution of LMXBs has already been modelled
by \citeauthor{ps88} (\citeyear{ps88}, \citeyear{ps89}), who predict
that binaries with starting periods $P_B\,\geq\,P_{\rm
bif}\,\simeq\,2$\,days increase their orbital separation, hence their
orbital period, during the mass transfer (diverging systems), while
the ones with $P_B\,\leq\,P_{\rm bif}$ decrease their orbital
separation (converging systems). In this second case, a simple argument
shows that the binary period decreases if $|\dot{M}_{\rm tot}|/M_{\rm
tot}\,<\,|\dot{a}|/a$ and increases if $|\dot{M}_{\rm tot}|/M_{\rm
tot}\,>\,|\dot{a}|/a$, where $M_{\rm tot}$ is the total mass of the
binary and $a$ is its orbital separation.

TS99 focussed their attention on diverging LMXBs and they
obtain a positive correlation between the orbital period and the final
mass of the HeWD companion, i.e., its mass increases with the
observed orbital period, and a negative correlation between the
orbital period and the mass of the pulsar. In order to compare the
theoretical results of TS99 to the observed systems, we searched the
literature for low-mass binary pulsars for which a reliable
measurement of both the pulsar and the companion mass has been
obtained. These values are reported in Table\,\ref{tab:psrmasses}
jointly with the binary period and the method used to obtain the
masses.

The masses of the two orbiting objects have been determined using
Shapiro delay for six binaries out of the 11 reported in
Table\,\ref{tab:psrmasses} (indicated with the label ``{\it r,s}'').
As we already discussed in \S\ref{subsec:impl}, this kind of
measurement does not need any a priori hypothesis, hence these six
objects can be used to perform six firm and independent tests on the
proposed models. The masses for the
remaining objects have been determined as follows:

\begin{itemize}
\item[i)] PSR\,J1012+5307 \citep{vbk96} and PSR\,B1957+20
\citep{vbk11}. The binary companion has been observed in phase
resolved spectroscopy observations in the optical band. This observing
mode leads to the determination of the companion mass function. Pulsar
timing gives the pulsar mass function and the knowledge of these two
quantities leads to the determination of the mass ratio of the system.
The companion mass has been determined by comparing the observed
spectroscopic properties to theoretical models for the atmosphere of HeWDs.
 \item[ii)] PSRs J1853+1303, J1910+1256 and J2016+1948
\citep{gsf+11}. The timing analysis gives a measurement of the proper
motion and of the first derivative of the pulsar projected semi-major
axis ($d(a_P\sin{\rm i})/dt$). This second quantity changes because
of the changing orbital inclination due to proper motion
(\citealt{kop96}, \citealt{ajrt96}). Since this also depends on the
unknown angle between the ascending node and proper motion directions,
the most probable orbital inclination has been determined using Monte
Carlo simulations. The companion mass is assumed to be that predicted
by the TS99 calculation, while the pulsar mass has been obtained by
combining the companion mass with the statistically determined orbital
inclination and the pulsar mass function.

\end{itemize}

\begin{figure} \centering
\includegraphics[angle=-90,width=15cm]{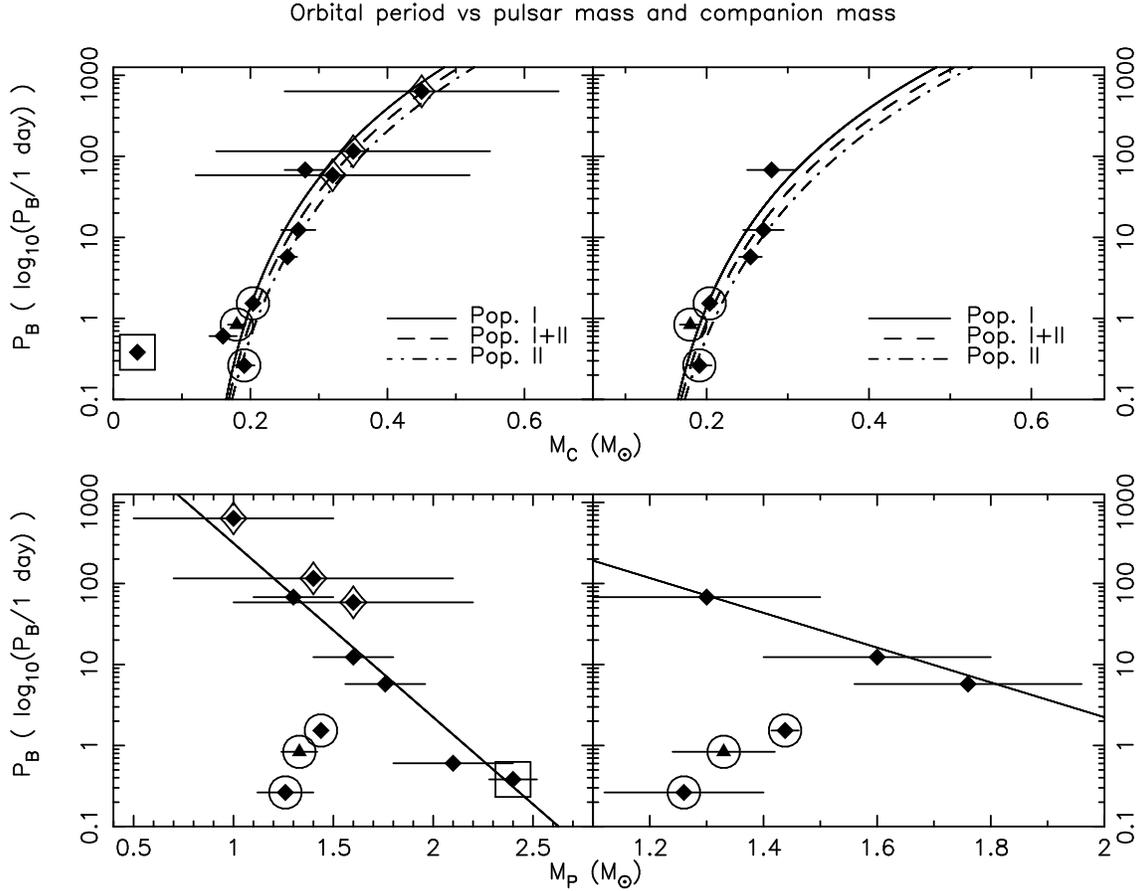}
\caption{ Orbital period versus companion mass (upper panels) and
  pulsar mass (lower panels) plots. The left panels show the points
  related to all binaries in Table\,\ref{tab:psrmasses}, while the
  right panels show the points that represent the binary systems for
  which the masses have been determined by detecting the Shapiro delay
  effect only. Error bars are $1\sigma$ uncertainties in the
  masses. If not visible, their size is smaller than the symbol. In
  all panels, the triangle represents PSRA (the only globular cluster
  object) diamonds represent field objects, the point enclosed by a
  square represents PSR\,B1957+20, the points enclosed by a diamond
  represent the three objects for which the determination of the
  companion mass is based on a model by TS99, and the circled points
  indicate the binaries that do not obey the linear in $\log$
  $M_P-P_B$ correlation (the solid line in the bottom plots). In the
  upper panels the solid, dashed and dotted lines represent the
  numerical results by TS99 (fig.4 in TS99 and text for details),
  while in the lower panels the solid lines represent our fit to all
  plotted points in the bottom left panel except the circled ones.
\label{fig:mpmcpb}
}
\end{figure}

In Fig.\,\ref{fig:mpmcpb} we plot the logarithm of the binary
period versus companion mass (upper panels) and the pulsar mass (lower
panels). We plotted all objects reported in
Table\,\ref{tab:psrmasses} in the left panels, while in the right ones
we only show those binaries for which the masses have been
determined from measured Shapiro delays. We marked with a triangle
the position of PSRA in these plots to distinguish it from the field
systems which are denoted by diamonds.

In the left upper panel of Fig.\,\ref{fig:mpmcpb} the solid, dashed
and dotted lines represent the predictions of the $M_C-P_B$
correlation by TS99 (equations\,20 and 21). With the exception of the
squared point in the left bottom part of the plot, all other points
well agree with the theoretical lines by TS99.  Intriguingly the
extrapolations of the results of TS99 to binary periods lower than
2\,days are also in agreement with the observations of binaries with
$P_B\,\leq\,2$\,days, i.e. systems whose progenitors were converging
LMXBs. This means that the hypothesis assumed by TS99 to model the
diverging LMXBs can predict rather well the final HeWD mass also for
converging systems. The three points highlighted by a diamond shape
represent the binaries for which the companion mass was obtained from
one of the TS99 models \citep{gsf+11}, so they of course fall on that
line in the $M_C-P_B$ plot. The squared point represents the position of
PSR\,B1957+20, a black widow system. For this latter system we cannot
compare the actual companion mass with the predictions by TS99,
because of the mass loss the companion underwent after the formation
of the MSP+WD system. Nevertheless for the binary period of this
system TS99 predicts a companion mass $\sim0.18\,M_\odot$, i.e., higher
than the observed one. This is consistent with the hypothesis that this
system obeyed the TS99 $M_C-P_B$ correlation at the epoch of the
formation of the MSP+WD binary.

We note that the overall agreement of the predicted $M_C-P_B$
correlation with the available measurements has been already discussed
by various authors (e.g. \citealt{vbjj05} and \citealt{l08}). However
these comparisons where based on a variety of (often) indirect methods
for determining  $M_C$, in many cases with rather large confidence
intervals for the value for $M_C$. In the upper right panel of
Fig.\,\ref{fig:mpmcpb} we display a test for the $M_C-P_B$ correlation
entirely based on measures obtained via the detection of the Shapiro
delay. The theoretical models for the $M_C-P_B$ correlation by TS99
result in agreement to all 6 test binaries of our sample including
PSRA, the only binary likely associated with a globular cluster.

In the lower left plot of Fig.\,\ref{fig:mpmcpb}, eight of 11 objects
show a linear correlation between $M_P$ and $P_B$, while the remaining
three, highlighted with a circle, belong to the group of six binaries for
which the masses have been determined via the Shapiro delay
detection. A comparison between this panel and Fig.\,4-b in TS99
indicates that the theoretical predictions overestimate by about
$0.5\,M_\odot$ the pulsar mass for a given orbital period in the
entire range considered by TS99. If a real correlation exists between
$M_P$ and $P_B$ for this class of binaries, it is not the one
predicted by TS99 as these authors already pointed out in their
comparison with the observed systems\footnote{The fact that the three
objects in \citet{gsf+11} obey the observed $M_P-P_B$ correlation can
be seen as an indirect test of the results in TS99 for the
$M_C-P_B$ correlation, since the predictions by TS99 have been used
only to obtain the companion mass.}.

The solid line is a fit for all plotted points but the circled
ones modelling the $M_P-P_B$ correlation by using the
following function:

\begin{equation}
\frac{M_P}{M_\odot}\,=\,A+B\log_{10}\left(\frac{P_B}{{\rm
1\,day}}\right) 
\end{equation}

\noindent and we obtain $A\,=\,2.263\pm0.005$ and
$B\,=\,-0.466\pm0.003$ ($\chi^2\,=\,0.91$ with 6 degrees of freedom).

In the bottom right panel of Fig.\,\ref{fig:mpmcpb} the 6 binaries
which represent our firm tests for TS99 are equally divided into two
groups: three of them obey the $M_P-P_B$ correlation mentioned above,
while the other three do not. The binaries in this second group also
have binary periods lower than the bifurcation period $P_{\rm
bif}\,\simeq\,2$\,days theoretically found by \citeauthor{ps88}
(\citeyear{ps88}, \citeyear{ps89}), hence their progenitors were
converging LMXBs, while the other three have binary periods higher
than $P_{\rm bif}$, so it is more likely that their progenitors
where diverging LMXBs.

The quantitative discrepancy between the TS99 predictions and the
observed systems in the $M_P-P_B$ diagram, combined with the
agreement in the $M_C-P_B$ diagram, indicates that in TS99 the amount
of mass released by the companion is well predicted, but the amount
of matter captured by the neutron star is overestimated.  This
means that the discrepancies between the theoretical and observed
$M_P-P_B$ diagrams result from an incorrect modelling by TS99 of
the mechanisms (one or more) that prevent the neutron star from capturing
all the matter released by its companion and are responsible for the
mass loss that the binary system undergoes in its LMXB phase. Two
mechanisms have been proposed, namely
the radio ejection model \citep{bpd+01} and the propeller model
\citep{t12}. Since our test binaries divide themselves in two separate
groups, which in turn should have experienced different evolutionary
paths for their LMXB progenitors, it is tempting to suggest that
the mechanism responsible for the unaccreted matter is different in
the two cases of converging and diverging LMXBs, or that in one case
only one mechanism takes place at some point of the evolution, while
in the other case both mechanisms play a non-negligible role.

The position in the $M_P-P_B$ plot (bottom left) of PSR\,J1012+5307
and PSR\,B1957+20 is apparently at odds with this picture. Their
orbital periods indicate that their progenitors where converging
LMXBs, but they agree with the possible $M_P-P_B$ correlation for
binaries whose progenitors were diverging LMXBs. A first possibility
is that the pulsar mass in these systems is overestimated. In fact, as
reported above, $M_P$ was obtained using indirect methods relying on
modelling of the outer layers of the companion star. If the values for
$M_P$ of PSR\,J1012+5307 and PSR\,B1957+20 are confirmed by future
observations, we could speculate that the family of converging LMXBs
must be in turn subdivided in two groups. For the first group the
mechanism(s) preventing the accretion of mass are similar to the
diverging systems and the resulting HeWD-MSP binaries obey the same
$M_P-P_B$ correlation.  The second group is instead characterised by
peculiar conditions that make the loss of matter from the binary more
efficient. Detailed numerical simulations are required to clarify this
point.

\section{SUMMARY}
\label{sec:summary}

In this work we have presented a timing solution for the binary
millisecond pulsar PSR\,J1910-5959A based on Parkes timing data with a
data span of 12 years.  Our new solution gives improved values for the
astrometric and rotational parameters compared to those reported in
papers I\&II, and contains a new description of the orbital motion
including the first detection of the relativistic Shapiro delay for a
low-mass millisecond pulsar in a globular cluster.  The measurement of
this effect allowed us to determine reliable masses for the two
orbiting stars. We have compared the values for the mass and the
radius of the companion to the theoretical mass-radius relationships
obtained for different chemical compositions, but the current
uncertainties on these parameters prevent us from firmly establishing
whether PSR\,J1910-5959A belongs to NGC\,6752 or not.  Finally we
compared the numerical results on the evolution of low-mass X-ray
binary systems obtained by TS99, which predict the presence of a
correlation between $M_C$ and $P_B$ and between $M_P$ and $P_B$ for
binary systems where a millisecond pulsar orbits a low-mass helium
white dwarf companion, to the sample of these objects for which the
masses of the two stars have been measured using Shapiro
delay. Observations confirm the $M_C-P_B$ correlation obtained by
TS99, and show a possible $M_P-P_B$ correlation for a subset of the
sample which is different to the one obtained by TS99. The numerical
calculations by these authors overestimate the pulsar mass by about
$0.5\,M_\odot$. These results confirm that the amount of mass lost by
the binary during the evolution in LMXB is always larger than
predicted by TS99.  Moreover they suggest that the low-mass binary
pulsars with a WD companion can be split in two groups, characterised
by significantly different efficiencies in the mechanisms responsible
for the loss of matter during their evolution in the LMXB phase.

\acknowledgements{\small A. C., A. P., and N. D. acknowledge support
for this research provided by INAF, under the national research grant
PRIN-INAF 2010, awarded with DP 28/2011. The Parkes radio telescope is
part of the Australia Telescope, which is funded by the Commonwealth
of Australia for operation as a National Facility managed by CSIRO.
We thank colleagues who assisted with the observations discussed in
this paper.}

\newpage 

\begin{deluxetable}{ll}
\tabletypesize{\footnotesize}
\tablecaption{Measured and derived parameters for PSR\,J1910-5959A.\label{TabTim}}
\tablewidth{0pt}
\tablehead{
\colhead{Parameter} & \colhead{PSR\,J1910-5959A}
}
\startdata
RA\,(J2000) & 19:11:42.75562(3) \\
DEC\,(J2000)&$-$ 59:58:26.9029(3)\\
$\mu_{\alpha}\cos\delta$ (mas yr$^{-1}$) & $-$3.08(6) \\
$\mu_{\delta}$ (mas yr$^{-1}$) & $-$3.97(6) \\
$P$\,(ms) & 3.26618657079054(9) \\
$\dot{P}$ ($10^{-21}$) & 2.94703(14) \\
DM\,(pc~cm$^{-3}$) & 33.6998(16) \\
DM1\,(pc~cm$^{-3}$~s$^{-1}$) &0.00136(36) \\
Binary model & DD \\
$P_{orb}$ (days) & 0.83711347691(3) \\
$a_P\sin i$\,(lt-s)  & 1.2060418(7) \\
$T_{0}$\,(MJD) & 51919.206480057(55) \\
$M_{c}^{(a)}$\,(${\rm M_\odot}$) & 0.180(18) \\
$\sin i^{(a)}$ & $\geq$0.9994 \\
$i^{(a)}$ (deg) & $\geq$88$^\circ$ \\
$f\left(M_{\rm PSR}\right)$ (${\rm M_\odot}$) & 0.00268782589(5) \\
$\mu$ (mas yr$^{-1}$) & 5.02(6) \\
Position Angle (deg) & 217.9(7) \\
Ref. Epoch (MJD) & 51920.000\\
MJD range & 51710$-$55911 \\
Number of TOAs & 1003 \\
Rms residuals ($\mu$s) & 5.3 \\
\enddata
\footnotesize{
Units of right ascension are hours, minutes, and seconds, and units of
declination are degrees, arcminutes, and arcseconds. Numbers enclosed
in parenthesis represent the $1\sigma$ uncertainties on the last
significant digit for all parameters, either measured and derived,
presented in this table. The values, and relative uncertainties, for
the parameters that describe the Shapiro delay (indicated with the
apex $a$) have been obtained through a Bayesian analysis of our timing
results. The proper motion's position angle is measured
counterclockwise starting from the north direction.}
\end{deluxetable}

\begin{deluxetable}{cccccc}
\tabletypesize{\footnotesize}
\rotate
\tablecaption{Masses for binary pulsars with a low mass white dwarf companion.
\label{tab:psrmasses}}
\tablewidth{0pt}
\tablehead{
\colhead{PSR} & \colhead{Pulsar mass ($M_\odot$)} &
\colhead{Companio mass ($M_\odot$)} &
\colhead{Orbital period (days)} &
\colhead{Method} &
\colhead{Reference }
}
\startdata
J0437-4715 & $1.76\pm0.20$ & $0.254\pm0.014$ & 5.74 & r,s & \citet{vbv+08} \\
J0751+1807 & $1.26\pm0.14$ & $0.191\pm0.015$ & 0.26 & r,s & \citet{nsk08} \\
J1012+5307 & $2.1\pm0.3$ & $0.16\pm0.02$ & 0.60 & optical & \citet{vbk96} \\
J1713+0747 & $1.3\pm0.2$ & $0.28\pm0.03$ & 67.83 & r,s & \citet{sns+05} \\
J1853+1303 & $1.4\pm0.7$ & $0.33-0.37$ & 115.65 & TS99 & \citet{gsf+11} \\
B1855+09 & $1.6\pm0.2$ & $0.270\pm0.025$ & 12.33 & r,s & \citet{s04} \\
J1909-3744 & $1.438\pm0.024$ & $0.2038\pm0.0022$ & 1.53 & r,s & \citet{jhb+05} \\
J1910+1256 & $1.6\pm0.6$ & $0.30-0.34$ & 58.47 & TS99 & \citet{gsf+11} \\
J1910-5959A & $1.33\pm0.11$& $0.180\pm0.018$ & 0.84 & r,s & this work \\
B1957+20 & $2.40\pm0.12$ & $0.035\pm0.002$ & 0.38 & optical & \citet{vbk11} \\
J2016+1948 & $1.0\pm0.5$ & $0.43-0.47$ & 635.04 & TS99 & \citet{gsf+11} \\
\enddata
\footnotesize{
Uncertainties are at the 1$\sigma$ level. Pulsars are listed in right
ascension order. In the column ``Method'' ``r,s'' means that the
masses have been measured by detecting the Shapiro delay, `` optical''
by analysing optical observations of the companion, TS99 by using the
numerical results by \citet{ts99a}. Orbital periods have been obtained
by browsing the {\sc psrcat} pulsar catalogue
(http://www.atnf.csiro.au/research/pulsar/psrcat, \citealt{mhth05}),
and they have been rounded to the second decimal digit since their
uncertainties are so small that they can be considered negligible in
the discussion in \S\ref{sec:corr}.
}
\end{deluxetable}


\end{document}